\documentclass[12pt]{article}
\usepackage{color}
\usepackage{float}
\usepackage{graphics,epsfig}
\usepackage{bm}
\usepackage{slashed}
\usepackage{graphicx}
\usepackage{amsmath}
\usepackage{amsfonts}
\usepackage{amssymb}
\usepackage{color}
\usepackage{dcolumn}

\numberwithin{equation}{section}

\begin{document}

\title{\textbf{Thermal phase transitions of dimensionally continued AdS black holes}}
\author{Xiao-Mei Kuang \thanks{xmeikuang@gmail.com} \ and Olivera Miskovic\thanks{olivera.miskovic@pucv.cl}\bigskip \\
{\small Instituto de F\'\i sica, Pontificia Universidad Cat\'olica de Valpara\'\i so,}\\
{\small Casilla 4059, Valpara\'{\i}so, Chile}}
\maketitle

\begin{abstract}
We study the thermal phase transitions of charged black holes in
dimensionally continued gravity in anti-de Sitter space. We find the van der
Waals-like phase transition in the temperature-entropy plane of the black holes with
spherical horizons in even dimensions, and there is no such phase transition of the black holes with flat and hyperbolic
geometries. Near the critical inflection point, the critical
exponent is computed and its value does not depend on the dimension. The Maxwell
equal area law is checked to be fulfilled in the temperature-entropy diagram
for the first order phase transition. In odd dimensions, there are no thermal
phase transitions for the black hole with any geometry of the horizon.
\end{abstract}

\section{Introduction}

The black hole thermodynamics is a challenging topic in black hole physics
that has been studied for decades because of its non-perturbative, quantum nature.
The phase diagrams of black holes exhibit rich structures that carry information about stability of these systems and, for that reason, have
been attracting more attention and effort with time. Among them, the van der Waals-like behavior of black holes has been widely
studied because of its striking similarity to the liquid-gas phase transition in the ordinary
thermodynamic systems.

The earliest analysis of first order phase transitions in black hole
systems can be found in refs.\cite{Chamblin:1999tk,Chamblin:1999hg},
where the authors studied the inverse temperature-horizon  diagram.
It was found that, as the charge of black hole decreases to a
critical value, the system undergoes a second-order phase transition, and
below the critical charge, the black hole always presents first-order phase
transition. This process is very similar to the one of the van der Waals liquid-gas
transition. Later, by treating the negative cosmological constant as
the pressure and the physical volume enclosed by the black hole horizon
as its thermodynamical conjugated quantity,
and identifying the black hole mass as enthalpy instead of internal energy,
the black hole thermodynamic is re-studied in the extended phase space
that includes the pressure and volume \cite{Kastor:2009wy,Cvetic:2010jb,Dolan1,Dolan2},
and the first law agrees with the corresponding Smarr relation
\cite{Rasheed:1997ns,Breton:2004qa,Huan:2010}. Further, first attempts to
study the phase structure in the pressure-volume, $P$-$V$, diagram was made
for the charged Reissner-N\"{o}rdstrom (RN) anti-de Sitter (AdS) black hole with
fixed charge and the van der Waals-like phase transition was observed  in
\cite{Kubiznak:2012wp}. Namely, as the temperature increases, the black hole
undergoes the first-order phase transition and then the second-order phase
transition at the critical point, until it arrives at a stable phase. This
proposal led to many generalizations in more general cases
\cite{Zou:2013owa,Mo:2014qsa,Hennigar:2015esa,Xu:2015rfa,Hendi:2015xya,Majhi:2016txt}.
In particular, in Ref. \cite{Majhi:2016txt}, the $PV$ criticality was discussed in a general framework,
based on the Smarr formula and the first law of thermodynamics, revealing its universality.

In this work, we are interested in the phase structure of charged black
holes in dimensionally continued gravity proposed in \cite{BTZ}. It is a
higher-dimensional Lovelock AdS theory which possesses a unique AdS vacuum
\cite{Crisostomo-Troncoso-Zanelli}, so that the only parameters in the
theory are the gravitational constant, $\kappa $, and the AdS radius, $\ell $.
In particular, this theory becomes Chern-Simons AdS gravity in odd dimensions
\cite{Chamseddine} and Born-Infeld AdS gravity in even dimensions. The
theory possesses topological black hole solutions whose properties and thermodynamics were
studied in \cite{BTZ,Cai:1998vy,Muniain:1995ih}. A full boundary series of
counterterms that renders dimensionally continued gravity finite in the IR
region, is presented in Ref.\cite{Miskovic-Olea}.

In this work  we will focus on thermal phase transitions in the
temperature-entropy, $T$-$S$, plane in ordinary black hole phase space,
rather than in the $P$-$V$ plane in the extended space, for the following
two reasons. On one hand, it was argued in \cite{Spallucci:2013osa} that
these two behaviors are dual to each other because  both of
them stem from the same equation, i.e., the expression for the Hawking
temperature of the black hole.
However, working in one side of the duality might be significantly simpler.
For example, the Maxwell's equal area law for charged AdS black holes was possible to solve exactly in the $T$-$S$ plane,
whereas the solution was found only perturbatively around the critical values in the P-V plane \cite{Spallucci:2013osa}. 
The van der Waals-like phase transition in the
$T$-$S$ plane was also observed in \cite{Zeng:2015tfj} and the references therein.

On the other hand, differently than in general relativity, the entropy of
dimensionally continued black holes depends on both the horizon $r_+$
and the cosmological constant through $\ell $. This leads
to a highly non-linear expression for the pressure, making it
technically very difficult to study phase diagrams in the extended phase space
of such non-linear objects. In fact, it is much more convenient
to analyze the phase structure in the $T$-$S$ plane, where the parameter
$\ell $ does not vary, expecting that its phase structure is similar to
the one in the $P$-$V$ plane due to the duality.

Let us also comment that, as nicely explained in a footnote of Ref.\cite{Spallucci:2013osa},
this conjectured duality of descriptions in two different planes is analogous to the T-duality of string theory,
where the spectrum of a closed string wrapped around a compact dimension of some radius $R$
can be seen as the spectrum of a closed string wrapped around a compact dimension of the inverse radius, $1/R$.
The coupling constants in the corresponding dual theories are inverse to each other. Similarly, in black hole thermodynamics,
the state equation can be understood either as the Hawking temperature $T$ for the black hole in a charged background,
or as the pressure $P$ of a van der Waals fluid enclosed in a (specific) volume $V$ and heated at the temperature $T$.
Considering the former treatment means to focus on the $T$-$S$ phase space, while the latter treatment means to focus on the $P$-$V$ phase space.
Exchanging the roles of $P$ and $T$ is analogous to exchanging Kaluza-Klein momentum and winding modes in string theory.
An inversion of a coupling constant due to the duality in black hole thermodynamics refers to the Boltzman constant
($k_{{\rm B}} \rightarrow 1/k_{{\rm B}}$).

In Dimensionally Continued Gravity, we find that the van der Waals-like phase transition only occurs in black
holes with spherical horizons in Born-Infeld AdS gravity in even
dimensions, and there are no such phase transitions in Chern-Simons AdS
gravity in odd dimensions. For the black holes that exhibit a van der Waals-like
behavior, our results show that the critical exponent near the
critical point is consistent with that found in the mean field theory and the
Maxwell equal area law is satisfied for the first order phase transition.

The rest of this paper is organized as follows. In Section \ref{Sec-backround},
we briefly review dimensionally continued gravity and
the charged, topological black holes. We discuss the existence of thermal phase transitions
and their properties, such as critical exponents and the Maxwell equal area
law, in even dimensions in Section \ref{Sec-even-PT} and odd dimensions in
Section \ref{Sec-odd-PT}, respectively. Section \ref{Sec-conclusion}
contains our conclusions and discussion.

\section{Charged dimensionally continued black holes}

\label{Sec-backround}

Lovelock gravity action is a linear combination of dimensionally continued
Euler densities \cite{love,Lanczos},
\begin{equation}
I_{G}=\kappa \int d^{D}x\sqrt{-g}\sum_{p=0}^{n-1}\frac{\alpha _{p}}{2^{p}}%
\,\delta _{\nu _{1}\cdots \nu _{2p}}^{\mu _{1}\cdots \mu _{2p}}\,R_{\mu
_{1}\mu _{2}}^{\nu _{1}\nu _{2}}\cdots R_{\mu _{2p-1}\mu _{2p}}^{\nu
_{2p-1}\nu _{2p}}\,,  \label{action}
\end{equation}
where an integer $n$ is related to the space-time dimension, so that $D=2n$ in even
dimensions and $D=2n-1$ in odd dimensions, and $\kappa $ describes the
strength of gravitational interaction. Here, $\delta _{\nu _{1}\cdots \nu
_{2p}}^{\mu _{1}\cdots \mu _{2p}}=\det \left[ \delta _{\nu _{1}}^{\mu
_{1}}\delta _{\nu _{2}}^{\mu _{2}}\cdots \delta _{\nu _{2p}}^{\mu _{2p}}%
\right] $ is the completely antisymmetric product of $2p$ Kronecker's
deltas.
The gravitational action (\ref{action}) is the most general one leading
to the second order field equations in the metric field $g_{\mu\nu}(x)$.
In a special case of dimensionally continued gravity (DCG) \cite{BTZ},
the coefficients $\alpha _{p}$ are fixed so that the vacuum solution
corresponds to the constant curvature space, $R_{\alpha \beta }^{\mu \nu }=-%
\frac{1}{\ell ^{2}}\,\delta _{\alpha \beta }^{\mu \nu }$, with the
AdS radius $\ell $. In the notation of Ref.\cite{Crisostomo-Troncoso-Zanelli},
the coefficients are\footnote{
In this reference, the action $I_{G}=\kappa \int \sum_{p=0}^{n-1}\tilde{%
\alpha}_{p}L_{p}$ is written in terms of differential forms $L_{p}=\epsilon
_{a_{1}...a_{D}}\,R^{a_{1}a_{2}}\cdots
R^{a_{2p-1}a_{2p}}\,e^{a_{2p+1}}\cdots e^{a_{D}}$. The coefficients
$\tilde{\alpha}_{p}$ are related to $\alpha _{p}$ as $\alpha _{p}=-\tilde{%
\alpha}_{p}\left( D-2p\right) !$.}
\begin{equation}
\alpha _{p}=-{\binom{{n-1}}{p}}\frac{\left( D-2p-1\right) !}{\ell ^{2(n-p-1)}%
}\,.  \label{coeff}
\end{equation}
From (\ref{action}) and (\ref{coeff}), it is clear that the only parameters in the theory are $\kappa$ and $\ell$.
To obtain charged solutions, one has to couple gravity to the
electromagnetic field,
\begin{equation}
I=I_{G}-\frac{1}{4e^{2}}\,\int d^{D}x\,\sqrt{-g}\,F^{2}\,,
\end{equation}%
with $e^{2}$ controlling the strength of electromagnetic interaction.
Equations of motion obtained from the action principle are the Maxwell
equation,
\begin{equation}
\nabla _{\nu }F^{\mu \nu }=0\,, \label{M}
\end{equation}
and the gravitational equation for DCG,
\begin{equation}
-\kappa \sum_{p=0}^{n-1}\frac{\alpha _{p}}{2^{p-1}}\,\delta _{\mu \mu
_{1}\cdots \mu _{2p}}^{\nu \nu _{1}\cdots \nu _{2p}}\,R_{\nu _{1}\nu
_{2}}^{\mu _{1}\mu _{2}}\cdots R_{\nu _{2p-1}\nu _{2p}}^{\mu _{2p-1}\mu
_{2p}}=\frac{1}{e^{2}}\left( \frac{1}{2}\,\delta _{\mu }^{\nu
}\,F^{2}-2F_{\mu \lambda }F^{\nu \lambda }\right) \,, \label{eomDCG}
\end{equation}
or equivalently
\begin{eqnarray}
&&-\frac{\kappa }{2^{n-2}}\,\delta _{\mu \mu _{2}\cdots \mu _{2n-1}}^{\nu
\nu _{2}\cdots \nu _{2n-1}}\,\left( R_{\nu _{2}\nu _{3}}^{\mu _{2}\mu _{3}}+%
\frac{1}{\ell ^{2}}\,\delta _{\nu _{2}\nu _{3}}^{\mu _{2}\mu _{3}}\right)
\cdots \left( R_{\nu _{2n-2}\nu _{2n-1}}^{\mu _{2n-2}\mu _{2n-1}}+\frac{1}{%
\ell ^{2}}\,\delta _{\nu _{2n-2}\nu _{2n-1}}^{\mu _{2n-2}\mu _{2n-1}}\right)
\notag \\
&&\qquad \qquad \left. =\frac{1}{e^{2}}\left( \frac{1}{2}\,\delta _{\mu
}^{\nu }\,F^{2}-2F_{\mu \lambda }F^{\nu \lambda }\right) \,.\right.
\end{eqnarray}
From the above form of the gravitational equation it is clear that, in
absence of the matter fields, the equation $\left( R+\frac{1}{\ell
^{2}}\,\delta ^{\lbrack 2]}\right) ^{n-1}=0$ yields, as a particular solution,
a maximally symmetric space with the unique AdS radius $\ell$.

The field equations (\ref{M}) and (\ref{eomDCG}) possess a static, spherically symmetric solution. The
electromagnetic potential $A_{\mu }=\phi (r)\,\delta _{\mu }^{t}$ reads
\begin{equation}
\phi (r)=\phi _{\infty }-\frac{Q}{\left( D-3\right) r^{D-3}}\,,
\end{equation}%
where $Q$ is an electric charge of the black hole, with the electric field $%
F_{tr}=Q/r^{D-4}$. The metric describes topological black holes,%
\begin{equation}
ds^{2}=-f^{2}(r)\,dt^{2}+\frac{dr^{2}}{f^{2}(r)}+r^{2}d\Sigma _{D-2}^{2}\,,
\label{metric}
\end{equation}
where $d\Sigma _{D-2}^{2}=\gamma _{mn}(y)\,dy^{m}dy^{n}$ is a constant
curvature transversal section with spherical ($k=1$), hyperbolic ($k=-1$) or
planar ($k=0$) geometry, whose area is $\Omega _{D-2}$. First integral of
the equations of motion, with the integration constant $\mu $, is common for
all Lovelock gravities,%
\begin{equation}
\sum_{p=0}^{n-1}\frac{\alpha _{p}}{\left( D-2p-1\right) !}\left( \frac{%
k-f^{2}}{r^{2}}\right) ^{p}=\frac{\mu }{r^{D-3}}-\frac{Q^{2}}{\left(
D-3\right) \kappa e^{2}r^{2D-6}}\,,
\end{equation}%
and, in the DCG case, it becomes
\begin{equation}
\left( \frac{1}{\ell ^{2}}+\frac{k-f^{2}}{r^{2}}\right) ^{n-1}=\frac{\mu }{%
r^{D-1}}-\frac{Q^{2}}{\left( D-3\right) \kappa e^{2}r^{2D-4}}\,.
\end{equation}
It is convenient to normalize the gravitational and electromagnetic constants as
\begin{equation}
\kappa =\frac{1}{\Omega _{D-2}G}\,,\qquad e^{2}=\Omega _{D-2}\,.
\end{equation}%
Then the mass of the black hole $M$ and the integration constant $\mu $ are
related by \cite{Crisostomo-Troncoso-Zanelli}
\begin{equation}
\mu =2GM+\delta _{D,2n-1}\,,
\end{equation}%
where the additive constant is chosen so that the black hole horizon shrinks
to a point for $M\rightarrow 0$, and is non-vanishing only in Chern-Simons
gravity. The general solution becomes \cite{BTZ,Cai:1998vy}
\begin{equation}
f^{2}(r)=k+\frac{r^{2}}{\ell ^{2}}-\left( \frac{2GM+\delta _{D,2n-1}}{%
r^{D-2n+1}}-\frac{GQ^{2}}{\left( D-3\right) r^{2(D-n-1)}}\right) ^{\frac{1}{%
n-1}}\,.  \label{DCG BH}
\end{equation}
In what follows, we will also set $G=1$.

\section{Even-dimensional charged black holes and thermal phase transitions}
\label{Sec-even-PT}

In even dimensions $D=2n$, the spacetime of a charged topological black hole
 in DCG is described by the metric (\ref{metric}) with
the metric function (\ref{DCG BH}),
\begin{equation}
f^{2}(r)=k+\frac{r^{2}}{\ell ^{2}}-\left[ \frac{2M}{r}-\frac{Q^{2}}{%
(D-3)r^{D-2}}\right] ^{\frac{1}{n-1}}.  \label{g-a}
\end{equation}%
For $n=2$, the solution recovers the four-dimensional RN AdS black hole.

The equation of the horizon, $f^{2}(r_{+})=0$, enables to express the black
hole mass $M$ in terms of the horizon $r_{+}$ as
\begin{equation}
M=\frac{r_{+}}{2}\left[ \left( k+\frac{r_{+}^{2}}{\ell ^{2}}\right) ^{n-1}+%
\frac{Q^{2}}{(D-3)r_{+}^{D-2}}\right] \,.
\end{equation}
The black hole temperature $T$ is calculated in a standard way, from the Euclidean continuation of the spacetime,
where the Euclidean period $T^{-1}$ avoids a conical singularity near the horizon if $T=\frac{1}{4\pi}\,\partial_r f^{2}|_{r_{+}}$. Then the metric (\ref{g-a}) gives
\begin{equation}
T=\frac{1}{4\pi (n-1)r_{+}}\left[ k+(2n-1)\,\frac{r_{+}^{2}}{\ell ^{2}}-
\frac{Q^{2}}{r_{+}^{D-2}}\left( k+\frac{r_{+}^{2}}{\ell ^{2}}\right) ^{2-n}
\right] \,,  \label{T-a}
\end{equation}
and the entropy  has the form  \cite{Cai:1998vy}
\begin{equation}
S=\int^{r_{+}}_{0}dr_{+}\,\frac{1}{T}\left(\frac{\partial M}{\partial r_{+}}\right)_{Q}=\pi \ell ^{2}\left[ \left( k+\frac{r_{+}^{2}}{\ell ^{2}}\right) ^{n-1}-k%
\right] \,.  \label{entropy1}
\end{equation}
Thermal properties of these black holes have been discussed in \cite{Cai:1998vy,Muniain:1995ih}.
In these papers, the authors studied critical behavior of DCG black holes by identifying
the divergencies of thermal capacity, but no nature of these transitions was discussed. They
found that phase transitions are possible only in even dimensions.

In the current study, we are interested in the
van der Waals-like phase transitions in the $T$-$S$ plane. Namely, the
entopy (\ref{entropy1}) is a function of the horizon, thus from the state
equation (\ref{T-a}), the temperature --which is a function of the horizon
$r_{+}$ and the charge $Q$-- can be seen as a function of the entropy and
the charge, $T=T(S,Q)$. The $T$-$S$ diagram corresponds to a curve
in the $T$-$S$ plane when the charge is kept fixed.

First we focus on the case with compact horizon  ($k=1$) and work in the ensemble with the
fixed charge $Q$.

Combining (\ref{T-a}) and (\ref{entropy1}), we eliminate $r_{+}$ from the
equations and obtain the temperature $T$ in terms of the entropy $S$ expressed via the variable $s=\frac{S}{\pi\ell^2}+1$,
\begin{equation}
T=\frac{-\ell^{2-2n}\,Q^2\,s^{\frac{2-n}{n-1}}+\left((2n-1) s^{\frac{1}{n-1}}-2n+2\right)
\left( s^{\frac{1}{n-1}}-1\right) ^{n-1}}{4\pi\ell(n-1)\left( s^{\frac{1}{n-1}}-1\right) ^{n-\frac{1}{2}}}\,.  \label{T(S)}
\end{equation}
In first few even dimensions, the state equation $T(S,Q)$ reads
\begin{eqnarray}
n &=&2:\quad T=\frac{-Q^{2}+3\ell ^{2}s^{2}-5\ell ^{2}s+2\ell ^{2}}{4\pi
\ell ^{3}(s-1)^{3/2}}\,,  \nonumber \\
n &=&3:\quad T=\frac{-Q^{2}+5\ell ^{4}s^{2}-14\ell ^{4}s^{3/2}+13\ell
^{4}s-4\ell ^{4}\sqrt{s}}{8\pi \ell ^{5}\sqrt{s}\left( \sqrt{s}-1\right)
^{5/2}}\,,  \nonumber \\
n &=&4:\quad T=\frac{-Q^{2}+7\ell ^{6}s^{2}-27\ell ^{6}s^{5/3}+39\ell
^{6}s^{4/3}-25\ell ^{6}s+6\ell ^{6}s^{2/3}}{12\pi \ell
^{7}s^{2/3}(s^{1/3}-1)^{7/2}}\,.\label{T_S}
\end{eqnarray}
Similarly as in case of the van-der Waals $P$-$V$ diagram, the
critical point $(S_{c},Q_{c},T_{c})$ in the $T$-$S$ plane is obtained as the
inflection point of the curve $T(S)$ for constant $Q_c$ that satisfies the conditions
\begin{equation}
\left. \frac{\partial T}{\partial S}\right\vert _{Q}=0\,,\qquad \left. \frac{%
\partial ^{2}T}{\partial S^{2}}\right\vert _{Q}=0\,.  \label{dTdS}
\end{equation}
Requiring that first derivative of (\ref{T(S)}) with respect to the entropy
(or the variable $s$) vanishes, we find the electric charge as a function of
entropy,
\begin{equation}
Q^{2}_c=-\frac{\ell^{2n-2}(2n-1) s^{\frac{1}{n-1}}-2n}{\left( 4n-5\right)
\,s^{\frac{1}{n-1}}-2n+4}\,s\left( s^{\frac{1}{n-1}}-1\right) ^{n-1}\,,
\label{Q(S)}
\end{equation}
from where the temperature for critical $Q_c$ becomes
\begin{equation}
T_c=\frac{2(2n-1) \,s^{\frac{2}{n-1}}-\left( 6n-7\right) \,s^{%
\frac{1}{n-1}}+2(n-2)}{2\pi \ell \left( s^{\frac{1}{n-1}}-1\right) ^{%
\frac{1}{2}}\left[ \left( 4n-5\right) \,s^{\frac{1}{n-1}}-2\left( n-2\right) %
\right] }\,.  \label{Tc}
\end{equation}
Then, taking second derivative of (\ref{T(S)}) with respect to the entropy
and after that plugging in the solution for the charge (\ref{Q(S)}), we get
\begin{equation}
\left. \frac{\partial ^{2}T}{\partial S^{2}}\right\vert _{Q_c}
=\frac{\pi \,P_{n}(s^{\frac{1}{n-1}})}{4\ell(n-1)^{2}s^{\frac{2n-3}{n-1}}\left( \left(
4n-5\right) \,s^{\frac{1}{n-1}}+4-2n\right) \left( s^{\frac{1}{n-1}%
}-1\right) }\,,
\end{equation}%
where the polynomial $P_n(x)$ is always cubic in $x=s^{\frac{1}{n-1}}$, and is given by
\begin{equation}
P_{n}(x)=(2n-1)(4n-5) \,x^{3}+\left(-16n^{2}+29n-\frac{9}{2}\right)\,x^{2}+10n(n-2)\,x-2n (n-2) \,.
\end{equation}
In first few even dimensions, it has the form
\begin{eqnarray}
P_{2}(x) &=&9x^{3}-\frac{21}{2}\,x^{2}\,,  \notag \\
P_{3}(x) &=&35x^{3}-\frac{123}{2}\,x^{2}+30x-6\,,  \notag \\
P_{4}(x) &=&77x^{3}-\frac{289}{2}\,x^{2}+80x-16\,.
\end{eqnarray}
The inflection points are identified from $P_{n}=0$. We search for strictly
positive (real) solutions $x$. In four dimensions, for
example, we find
\begin{equation}
P_{2}=0\quad \Rightarrow \quad x=s_c=\frac{7}{6}\,,
\end{equation}
so that the critical value of the quantities obtained from (\ref{Q(S)}) and (%
\ref{T(S)}) are%
\begin{eqnarray}
S_{c} &=&\frac{\pi\ell^2 }{6}\simeq 0.5236\,\ell^2 \,,  \notag \\
Q_{c} &=&\frac{\ell }{6}\simeq 0.1667\,\ell\, ,  \notag \\
T_{c} &=&\frac{1}{\pi\ell}\sqrt{\frac{2}{3}}\simeq \frac{0.2599}{\ell}\,.
\end{eqnarray}

An asymptotically flat limit of spacetime corresponds to $\ell \rightarrow
\infty $. In that case, from (\ref{Tc}), the critical temperature vanishes,
because the entropy parameter $s$ is just a dimensionless number obtained as a root of $P_n$.
This means that the van del Waals-like critical behavior occurs only in an
spacetime possessing a cosmological constant, as it was observed in reference \cite{Kubiznak:2012wp}.

From now on, we shall set the AdS radius $\ell =1$, for the sake of simplicity.

The critical points exist in higher dimensions as well, but the polynomial
$P_{n}=0$ is more difficult to solve analytically when $n>2$. However, it can
be checked that $P_{n}$ has the discriminant zero ($n=2$) or negative ($n>2$), so there is only one real root and it is always positive, which means
that there is always exactly one critical point in each even dimension. We list the corresponding critical
points in various even dimensions in table \ref{table1}. We see that larger
dimension increases the critical entropy, while it suppresses the critical
charge and temperature.
\begin{table}[tbp]
\center{
\begin{tabular}{|c|c||c|c|c|}\hline
$D$&$n$&$S_c$&$Q_c$&$T_c$\\\hline
4&2&$0.5236$&$0.1667$&$0.2599$\\\hline
6&3&$0.9058$&$0.03583$&$0.1747$\\\hline
8&4&$1.1366$&$0.007146$&$0.1391$\\\hline
 \end{tabular}
\caption{\label{table1} The critical points with $k=1$ in different even dimensions.}}
\end{table}
\begin{figure}[h]
\centering
\includegraphics[width=0.9\textwidth]{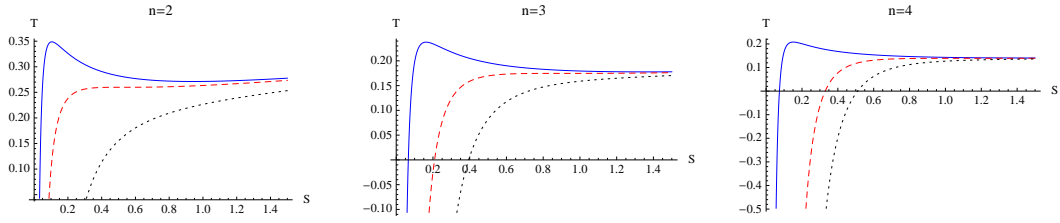}
\caption{$T$-$S$ diagram for the black hole with fixed charge in different even
dimensions. In each plot, the red dashed line corresponds to the related
critical charge $Q_{c}$, listed in table \protect\ref{table1}. The blue line and
black dotted line are related to the fixed charge which is smaller and
larger than $Q_{c}$, respectively. }
\label{fig-S-T}
\end{figure}
\begin{figure}[h]
\centering
\includegraphics[width=0.9\textwidth]{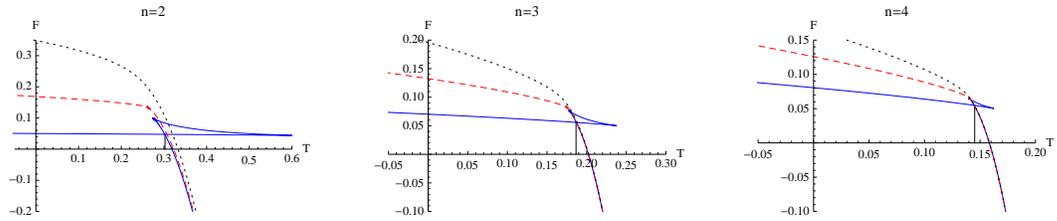}
\caption{Free energy in terms of the temperature. The temperature of the
crossing point in the blue line is the first order phase transition
temperature $T_{s}$ for $Q=0.3Q_{c}$. It shrinks into $T_{s}=T_{c}$ as $Q$
increases to $Q_{c}$, as the red dashed line shows. }
\label{fig-T-F}
\end{figure}

The $S$-$T$ diagrams related to the equation (\ref{T_S}) for different even
dimensions are shown in figure \ref{fig-S-T}. It is clear that each
plot possesses a boundary curve corresponding to the critical charge $Q_c=const$ (red dashed line), which shows an inflection point describing a second order phase transition  at the critical point.
 The specific heat capacity is divergent there. The lines below $Q_c$ (represented by a black dotted line), with $Q>Q_c$, are monotonous curves without inflection points,
 which means there is no phase transition in these cases.
The isocharge lines above $Q_c$ (represented by a blue line), with $Q<Q_c$, are the lines which are not monotonous, and their shape
is similar to the one describing a van der Waals phase transition.

More specifically, for the solid lines in each plot with fixed charge $Q<Q_{c}$,
there are three black holes sharing one temperature but with different free
energies to compete. There exists a critical temperature $T_{s}$ below which
the smallest black hole always has the lowest free energy and is
thermodynamically stable, and above which the largest black is stable. For
the dashed line with $Q=Q_{c}$, the two black holes merge into one at $%
T_{s}=T_{c}$. For the dotted line with $Q>Q_{c}$, there is no any
competition. This phenomenon is described in figure \ref{fig-T-F}, where we
show the relation between the temperature and free energy
\begin{equation}
F=M-TS\,.
\end{equation}

On the other hand, for a small charge with $Q=0.3\,Q_{c}<Q_c$ (blue lines), the
free energy graph shows a `swallow tail' in the $F$-$T$ plane, which is typical for
a first order phase transition. The horizontal coordinates of the black line
in each plot denote the first order phase transition temperature $T_{s}$
with the related parameters. Furthermore, for the critical charge $Q_c$ (red dashed lines)
at the inflection point where we have $T_{s}=T_{c}$, the curves show second order
phase transition because equation \eqref{fig-T-F} implies that the specific
heat capacity $C_{Q}=T\left. \frac{\partial S}{\partial T}\right\vert _{Q}$ is divergent.
Critical phenomena discussion from a divergence of $C_{Q}$ has also been developed in \cite{Banerjee:2011cz,Banerjee:2012zm}.

Near the critical inflection point $(Q_{c},S_{c},T_{c})$, we can calculate
the critical exponent $\alpha $, which characterizes a behavior of the heat capacity near the critical point as
\begin{equation}
C_{Q}=T\left. \frac{\partial S}{\partial T}\right\vert _{Q}\simeq
A(T-T_{c})^{\alpha }.
\end{equation}%
To determine $\alpha $, we expand the temperature \eqref{T_S} near the
critical point $S_{c}$, when $Q_{c}$ is kept fixed,
\begin{equation}
T=T(S_{c},Q_{c})+\left. \frac{\partial T}{\partial S}\right\vert
_{c}(S-S_{c})+\frac{1}{2}\left. \frac{\partial ^{2}T}{\partial S^{2}}\right\vert _{c}(S-S_{c})^{2}+\frac{1}{6}\left. \frac{\partial ^{3}T}{\partial S^{3}}\right\vert _{c}(S-S_{c})^{3}+\mathcal{O}((S-S_{c})^{4})\,.
\label{T-expansion}
\end{equation}%
Here\ `$c$' denotes the point $(Q_{c},S_{c})$. The first term is
$T(S_{c},Q_{c})=T_{c}$, while from the conditions \eqref{dTdS}, the second
and third terms of \eqref{T-expansion} vanish. This yields near the
critical point
\begin{equation}
T-T_{c}\simeq \frac{1}{6}\left. \frac{\partial ^{3}T}{\partial S^{3}}%
\right\vert _{c}(S-S_{c})^{3}\,.  \label{near c}
\end{equation}%
We also calculate $\left.\frac{\partial S}{\partial T}\right|_Q$ expanded near the critical point
by taking a partial derivative in $S$ of the expression (\ref{near c}),
\begin{equation}
\frac{\partial T}{\partial S}\simeq \frac{1}{2}\left. \frac{\partial ^{3}T}{%
\partial S^{3}}\right\vert _{c}(S-S_{c})^{2}\,\quad \Rightarrow \quad \frac{%
\partial S}{\partial T}\simeq 2\left. \frac{\partial ^{3}T}{\partial S^{3}}%
\right\vert _{c}^{-1}(S-S_{c})^{-2}\,.
\end{equation}%
Thus, the heat capacity expands as%
\begin{equation}
C_{Q}=T\left. \frac{\partial S}{\partial T}\right\vert _{Q}\simeq
2T_{c}\left. \frac{\partial ^{3}T}{\partial S^{3}}\right\vert
_{c}^{-1}(S-S_{c})^{-2}\,.
\end{equation}%
Now we invert the series (\ref{near c}),%
\begin{equation}
S-S_{c}\simeq \sqrt[3]{6}\left. \frac{\partial ^{3}T}{\partial S^{3}}%
\right\vert _{c}^{-\frac{1}{3}}\left( T-T_{c}\right) ^{\frac{1}{3}}\,,
\end{equation}%
and obtain%
\begin{equation}
C_{Q}\simeq A\,\left( T-T_{c}\right) ^{-\frac{2}{3}}\,.
\end{equation}%
We conclude that the heat capacity diverges at the critical point, as expected. The constant $%
A=\frac{1}{3}\sqrt[3]{6}T_{c}\left( \partial _{S}^{3}T\right) _{c}^{-1/3}$
can be evaluated explicitly by taking derivatives of (\ref{T(S)}). We
read-off the critical exponent $\alpha =-2/3$, which is independent of the
dimension. This exponent agrees with the one found in the mean field theory.
A general approach to obtain a critical exponent of black hole phase transition has been proposed in refs.\cite{Mandal:2016anc,Banerjee:2016nse}, where it was shown that the result does not depend on a particular black hole solution.

\begin{table}[tbp]
\center{
\begin{tabular}{|c|c||c|c|c|c|c|c|}\hline
$D$&$n$&$T_s$&$S_l$&$S_m$&$S_s$&$\eqref{Maxwell law}_{L}$&$\eqref{Maxwell law}_{R}$\\\hline
4&2&$0.3020$&$2.5405$&$0.4121$&$0.009865$&$0.7534$&$0.7533$\\\hline
6&3&$0.1870$&$2.8250$&$0.6703$&$0.09581$&$0.5104$&$0.5103$\\\hline
8&4&$0.1448$&$2.7920$&$0.8563$&$0.2300$&$0.3709$&$0.3708$\\\hline
\end{tabular}
\caption{\label{table2} The first order phase transition temperature with fixed $Q=0.3Q_c$ and  $k=1$ in different even dimensions, and the values of \eqref{Maxwell law} which show that
the Maxwell area law  is satisfied. $\eqref{Maxwell law}_L$ and $\eqref{Maxwell law}_R$ denote the values of the left and right sides of equation \eqref{Maxwell law}.}}
\end{table}

For the first order phase transition with $Q<Q_{c}$, the physical $T$-$S$
diagram should be modified by replacing the oscillating part by an isobar
with $T=T_{s}$, where $T_{s}$ is the first phase transition temperature.
Since, at the phase transition point, the free energy does not change,
integration of the first law of thermodynamics implies $\oint SdT=0$. This leads to
the Maxwell's equal area law
\begin{equation}
T_{s}(S_{l}-S_{s})=\int\limits_{S_{s}}^{S_{l}}TdS\,,  \label{Maxwell law}
\end{equation}%
where $S_{l}$ and $S_{s}$ are the largest and smallest entropies of
three intersection points, respectively, between the $S$-$T$ diagram and the
related isobar $T=T_{s}$, i.e., the three solutions $(S_{l},S_{m},S_{s})$ of
$S$ to the equation \eqref{T_S} with $T=T_{s}$ for fixed charge. We
calculated both sides of equation \eqref{Maxwell law} with $%
Q=0.3\,Q_{c}$, and the results are summarized in table \ref{table2}. It
turns out that the Maxwell equal area law \eqref{Maxwell law} is satisfied
in considered even dimensions, with an acceptable error between analytical (left
side) and numerical (right side) calculation.

So far, we have obtained that for the charged Born-Infeld black hole with
the spherical horizon ($k=1$), there exists a critical inflection point
$(Q_{c},S_{c},T_{c})$, near which the critical exponent matches the values in
the mean field theory. The system presents a van der Waals-like phase transition
in the $T$-$S$ plane of the state equation and the Maxwell equal area law is
checked to be satisfied.

In cases with non-compact horizons ($k=-1$ and $k=0$), the state equation (%
\ref{T(S)}) generalizes to
\begin{equation}
T=\frac{-Q^{2}\,s^{\frac{2-n}{n-1}}+\left( (2n-1)\,s^{\frac{1}{n-1}%
}-(2n-2)k\right) \left( s^{\frac{1}{n-1}}-k\right) ^{n-1}}{4\pi (n-1)\left(
s^{\frac{1}{n-1}}-k\right) ^{n-\frac{1}{2}}}\,,
\end{equation}%
where $s=\frac{S}{\pi }+k$. Using the same analysis as above, we write the
inflection point equations (\ref{dTdS}) and find that there are no positive
solutions for $S_{c}$ and $Q_{c}$. Explicitly, in four dimensions ($n=2$),
we get analytically%
\begin{eqnarray}
S_{c} &=&\frac{4k-3}{6}\,\pi \,,  \notag \\
Q_{c} &=&\frac{1}{6}\sqrt{52k^{2}-60k+9}\,,
\end{eqnarray}%
so the entropy becomes negative when $k=0$ or $-1$. Similar result
holds in higher dimensions, as well. This means that there are no van der Waals-like
phase transitions in black holes with non-compact horizons. This result that the existence of the van der Waals-like
phase depends on the topology of the horizon is also
valid in the extended phase space of RN AdS black holes \cite{Kubiznak:2012wp}.

\section{Odd-dimensional charged black holes and thermal phase transitions}

\label{Sec-odd-PT} In odd dimensions $D=2n-1$, the charged Chern-Simons AdS
black hole (\ref{DCG BH}) with $G=1$ has the form \cite{BTZ},
\begin{equation}
f^{2}(r)=k+\frac{r^{2}}{\ell ^{2}}-\left( 2M+1-\frac{Q^{2}}{(D-3)r^{D-3}}%
\right) ^{\frac{1}{n-1}}\,.  \label{g-b}
\end{equation}
In three dimensions ($n=2$), this solution is the BTZ black hole
\cite{BTZ-BH}.

The Hawking temperature for an arbitrary $n$ reads
\begin{equation}
T=\frac{r_{+}}{2\pi \ell ^{2}}-\frac{Q^{2}\left( k+\frac{r_{+}^{2}}{\ell ^{2}%
}\right) ^{2-n}}{4\pi (n-1)r_{+}^{2n-3}}\,,  \label{T-b}
\end{equation}%
and the entropy is given in the parametric form,%
\begin{equation}
S=4\pi (n-1)r_{+}\int\limits_{0}^{1}du\left( k+u^{2}\,\frac{r_{+}^{2}}{\ell
^{2}}\right) ^{n-2}.
\end{equation}

We apply the same strategy as in the last section. Since the equation
$S(r_{+})$ cannot be inverted in a simple way to obtain $r_{+}(S)$, it is
convenient to calculate derivatives in eqs.(\ref{dTdS}) as
\begin{equation}
\left. \frac{\partial T}{\partial S}\right\vert _{Q}=\frac{\left. \frac{%
\partial T}{\partial r_{+}}\right\vert _{Q}}{\frac{dS}{dr_{+}}}\,.
\end{equation}%
We find (with $\ell =1$)%
\begin{eqnarray}
\frac{dS}{dr_{+}} &=&4\pi (n-1)\int\limits_{0}^{1}du\left[ k+\left(
2n-3\right) \,u^{2}r_{+}^{2}\right] \left( k+u^{2}r_{+}^{2}\right) ^{n-3}
\notag \\
&=&4\pi (n-1)\left( k+r_{+}^{2}\right) ^{n-2},
\end{eqnarray}%
as well as%
\begin{equation}
\left. \frac{\partial T}{\partial r_{+}}\right\vert _{Q}=\frac{1}{2\pi }+%
\frac{Q^{2}\left[ \left( 2n-3\right) k+\left( 4n-7\right) r_{+}^{2}\right]
\left( k+r_{+}^{2}\right) ^{1-n}}{4\pi (n-1)r_{+}^{2n-2}}\,,
\end{equation}%
what implies vanishing $\left. \partial T/\partial S\right\vert _{Q}$ for%
\begin{equation}
Q^{2}_c=-\frac{2(n-1)r_{+}^{2n-2}}{\left[ \left( 2n-3\right) k+\left(
4n-7\right) r_{+}^{2}\right] \left( k+r_{+}^{2}\right) ^{1-n}}\,,\quad
r_{+}\neq -k\text{.}  \label{Q2}
\end{equation}%
It is clear that $Q^{2}_c$ can be positive only for hyperbolic black holes, so
the planar ones are ruled out of having a phase transition of the considered
type. Setting $k=-1$ and using again the same method to calculate the second
derivative of the entropy, we obtain that
$\left. \frac{\partial ^{2}T}{\partial S^{2}}\right\vert_Q$
replaced from (\ref{Q2})) vanishes when
\begin{equation}
\left( 8n^{2}-26n+21\right) r_{+}^{4}-\left( 8n^{2}-25n+20\right)
r_{+}^{2}+2n^{2}-5n+3=0\,.
\end{equation}%
The discriminant of this quadratic polynomial in $r_+^2$ is always negative when $n>2$, so there is
no a real solution for $r_{+}$ and, therefore, there is no a critical
point. When $n=2$, the root is $r_{+}=1$, but this point is not allowed
because of the inequality in (\ref{Q2}). An independent analysis in three
dimensions shows that $Q_{c}^{2}$ becomes negative, so again there is no
a critical point when $n=2$.

We conclude that, in all odd dimensions, Chern-Simons AdS topological black
holes, including the BTZ black hole, do not admit a van der Waals-like phase
transition for any geometry of the horizon.

\section{Conclusions and discussion}

\label{Sec-conclusion}

We analyzed the thermal phase transitions of
charged dimensionally continued black holes in the  $T$-$S$ plane.
In even dimensions, we found that the critical inflection point
$(Q_{c},S_{c},T_{c})$ can only exist if the black hole has a
spherical horizon (with $k=1$), and not if it has a non-compact geometry ($k=0$ or $k=-1$.)
For $k=1$, near the critical point, the
critical exponent in the specific heat capacity is always $-2/3$ in any even
dimension and this values agree with that of the mean field theory.
Moreover, we found that the system goes through a van der Waals-like
phase transition. When the charge is smaller than the critical charge, it
undergoes the first order phase transition at $T=T_{s}$, below which the
smallest black hole always has the lowest free energy and is thermodynamically
stable while, above it, the largest black hole is thermodynamically stable.
The Maxwell equal area law has been checked to be satisfied at the first
order phase transition point in first few even dimensions.

We did not observe a van der Waals-like phase transition in odd-dimensional
black holes in the charged dimensionally continued gravity.

It is important to mention that, at first sight, two families of DCG, Chern-Simons AdS and Born-Infeld
AdS gravity, seem similar, but they are intrinsically different. First,
Chern-Simons gravity comes from a Chern-Simons form, whose exterior
derivative is a topological invariant. In contrast, Born-Infeld gravity does
not have a geometric origin. Furthermore, in comparison to all Lovelock AdS
actions, Chern-Simons AdS features a symmetry enhancement from local Lorentz to
local AdS group, what drastically changes its dynamic structure and a number of
local degrees of freedom. While all generic Lovelock gravities have the same
number of degrees of freedom as General Relativity \cite{Teitelboim}, even a
Pure Lovelock \cite{Dadhich:2015ivt} which does not posses the linear in
curvature term, this is no longer true for Chern-Simons gravity \cite%
{Banados-Garay-Henneaux}. From this point of view, it is not surprising that
its phase space structure is drastically different than the one of
Born-Infeld.

It is worthwhile noticing that, with the development of holographic duality,
it was observed in \cite{Johnson:2013dka} that there also exists the van der
Waals-like phase transition in the entanglement entropy-temperature ($S_{E}$-$T$)
diagram by studying the holographic entanglement entropy (see \cite{Ryu:2006ef} for review)
in a finite volume quantum system dual to a charged
AdS black hole with spherical geometry. More related studies in
\cite{Zeng:2015tfj,Caceres:2015vsa,Dey:2015ytd,Sun:2016til,Zeng:2016aly} indicate
that similar to thermal entropy, the entanglement entropy also presents the
van der Waals-like phase behavior. Thus, it would be very interesting and
important to explore the holographic entanglement entropy of the charged
dimensional continued black holes, especially the one with spherical horizon
in even dimensions, in which the van der Waals-like phase transition in
$S$-$T$ has been observed. We will address this question in the near future.

\section*{Acknowledgments}

This work was funded by the Chilean FONDECYT Grant No.3150006 and the VRIEA-PUCV
grant N{o.}039.345/2016.

\end{document}